\documentclass[twocolumn,aps,prb,showpacs,preprintnumbers,amsmath,amssymb,psfrac]{revtex4}

\usepackage{epsfig}
\usepackage{graphicx}
\usepackage{dcolumn}
\usepackage{bm}

\renewcommand{\dag}{^{\dagger}}
\newcommand{\dl}{\partial_\ell}

\begin{document}
%

 \title{
Tomonaga-Luttinger model with an impurity for a weak two-body interaction
}
\author{Tobias Stauber}
\affiliation{Institut f\"ur Theoretische Physik, Ruprecht-Karls-Universit\"at Heidelberg, Philosophenweg 19, D-69120 Heidelberg, Germany
}

\begin{abstract}
The Tomonaga-Luttinger model with impurity is studied by means of flow equations for Hamiltonians. The system is formulated within collective density fluctuations but no use of the bosonization formula is made. The truncation scheme includes operators consisting of up to four fermion operators and is valid for small electron-electron interactions. In this regime, the exact expression for the anomalous dimension is recovered. Furthermore, we verify the phase diagram of Kane and Fisher also for intermediate impurity strength. The approach can be extended to more general one-body potentials.
\end{abstract}
%
%
%
%
\maketitle
\section{Introduction}
Ten years ago, Kane and Fisher showed in their influential work on the transmission through barriers in one-dimensional (1D) conductors, that at $T=0$ an arbitrarily weak barrier leads to perfect reflection at the impurity in the presence of repulsive two-body interaction whereas not even an arbitrarily large barrier can hinder the electrons from perfect transmission through the impurity if there is attractive two-body interaction.\cite{Kan92} Employing the bosonic representation of a 1D fermionic system, the results were obtained from a perturbative renormalization group (RG) analysis in the {\em impurity strength}, treating the weak coupling as well as the dual strong coupling regime. Arguing that the RG flow cannot cross the marginal fixed point line of zero interaction, the authors obtained their universal results independent of the impurity strength. The approach and result is closely connected to the perturbational RG treatment of a Brownian particle in a periodic potential \cite{Fis85} which experiences a phase transition from diffusive to localized behavior at some critical coupling strength, first observed by Schmid.\cite{Sch83}

In the following years many authors confirmed and also disproved the above results and the controversy was only resolved four years ago in favor of the initial results by finite-size refermionization at the exactly solvable Toulouse-point.\cite{Del98} The reason for the steady interest is the experimental relevance of the results since 1D transport properties are mainly determined by the transmission through the barrier and thus show algebraic behavior as a function of the temperature involving non-universal exponents, in accordance with Luttinger liquid theory.\cite{Kan92,Hal81} Furthermore, the spectral density of states at the impurity site shows algebraic suppression at the chemical potential, but the power-law is governed by a larger exponent than the bulk exponent of a translational invariant model. Recently, evidence was found that the conductance in GaAs/AlGaAs quantum wires behaves according to the theoretical predictions based on the bosonization technique.\cite{Rot00} But it was also pointed out that finite size effects might become important since the bosonic representation of the 1D Fermi gas is only applicable within an asymptotically small regime around the Fermi energy.\cite{Med02}

In this work, we want to apply a novel renormalization scheme to the problem, the flow equation approach for Hamiltonians.\cite{Gla94,Weg94} The presented treatment turns out to be in a way complementary to the RG approach of Kane and Fisher as it is valid for small two-body interaction but {\em arbitrary} impurity strength. Further, it does not rely on the bosonization technique\cite{FootBos} and it can easily be extended to more general one-body potentials. The fact that the flow equation approach is a novel renormalization scheme involving different approximations can also be seen from the analysis of the two-dimensional Hubbard model. Whereas functional renormalization schemes\cite{Zan98} rely on the static limit $\omega=0$, this kind of approximation is not present in the flow equation approach since the Hamiltonian formulation of the problem is preserved. Unavoidable in the flow equation treatment is the neglect of higher order coupling terms in the on-site interaction $U$. Nevertheless, the anti-ferromagnetic instability of the strong coupling regime is predicted.\cite{Gro02} Similarly, we will - in the following - recover the {\em exact} coupling-dependent anomalous dimension even though our truncation scheme is valid only for small electron-electron interaction.
 
The paper is organized as follows. In Sec. II, we will first treat the Tomonaga-Luttinger (TL) model within the flow equation approach, deriving the exact results for correlation functions in the position-time space by making explicit use of the bosonization formula. We will then employ the usual representation of the fermionic field operator, where approximations become necessary. Keeping track of only the first non-trivial term, we will obtain the algebraic power-law behavior for {\em small} two-body interaction and the {\em exact} anomalous dimension is recovered. In Sec. III, we will add an impurity to the system and show within the same truncation scheme as in the case of the TL model that the backward scattering constant yields the dominant contribution and either tends to zero or infinity depending on (weak) attractive or repulsive interaction. Since the asymptotic behavior of the flow equations turns out to be independent of the  initial conditions, the result also holds for the regime of intermediate impurity strength which was not explicitly treated by the RG scheme of Kane and Fisher. We also evaluate the spectral density at the impurity after neglecting the irrelevant forward scattering term. In Sec. IV. we close with conclusions.

%
%
\section{Tomonaga-Luttinger Model}
\label{ExactTLM}

In the following section, we briefly derive the TL model starting from a continuous 1D interacting electron gas, mainly in order to introduce the notation. We then discuss the model via flow equations with and without the use of the bosonization formula. The TL model is closely related to the $n$-orbital model in 1D on which Wegner introduced the method of flow equations.\cite{Weg94} We also want to mention the recent work by Heidbrink and Uhrig on 1D spinless fermions on a lattice with nearest neighbor interaction using continuous unitary transformations.\cite{Hei02}

\subsection{The model} 

We start with the description of one-dimensional, non-relativistic, interacting spinless electrons on a ring of length $L$. The continuum version of the Hamiltonian reads
\begin{align}
  \label{TL-Hamiltonian}
  H&= \int \limits_{-L/2}^{L/2} dx \psi^{\dagger}(x)
  \left( -\frac{\partial_{x}^{\,2}}{2m} +U(x)\right) \psi(x)\\\notag
  &+\frac{1}{2}
  \int\limits_{-L/2}^{L/2}\int\limits_{-L/2}^{L/2} 
  dx\,dx'\psi^{\dagger}(x)\psi^{\dagger}(x')V(x-x') 
  \psi(x')\psi(x)
\end{align}
where the fermionic field $\psi(x)$ obeys periodic boundary conditions and the canonical anti-commutation relation. $V(x)$ resembles the two-body potential between the fermions and we also included a possible one-body potential $U(x)$, which shall be zero for this section. Performing the Fourier transformation $\psi(x)=L^{-1/2}\sum_ke^{ikx}c_k$, $V(x)=L^{-1}\sum_qe^{iqx}v_q$ with $k,q=2\pi n/L$, $n\in\mathbb{Z}$ and focusing on the low-energy properties in the high-density limit of the system, the energy dispersion can be linearized around the Fermi points $\pm k_F$ where $k_F=2\pi N/L$, $N$ being the number of electrons. With the Fermi velocity $v_F\equiv k_F/m$ the momentum representation of the Hamiltonian is then given by
\begin{align}
\label{Liniarized_Hamiltonian}
H=v_F\sum_k|k|c_k\dag c_k+\frac{1}{2L}\sum_{k,k'q}v_q c_{k-q}\dag c_k
c_{k'+q}\dag c_{k'}
\end{align}
where $c_k^{(\dagger)}$ creates (annihilates) the plane wave with wave number $k$. Further, we neglected - and will neglect throughout this work - contributions from the number-of-particle operator, since they vanish in the thermodynamic limit.

The system can be formulated by two anti-commuting species of fermions, the left- and right-movers, $c_k^{L/R}$, related to the two Fermi-points $\pm k_F$.\cite{Tom50} A mathematically more rigorous treatment is possible if the two branches of the linear dispersion are extended, ranging from $k=\infty$ to $k=-\infty$.\cite{Lut63} To assure a well-defined ground-state, the states with negative energy have to be filled up with Dirac fermions, $d_k$. After proper normalization, the creation and annihilation operators of the left- and right-moving density fluctuations then obey exact bosonic commutation relations. These operators are defined as 
\begin{align}
\notag
b_q^R\equiv n_q^{-1/2}\sum_kc_{k-q}^{R\dagger}c_k^R\quad,\quad
b_q^L\equiv n_q^{-1/2}\sum_kc_{k+q}^{L\dagger}c_k^L\quad, 
\end{align}
where $q>0$ and $n_q\equiv Lq/(2\pi)\in\mathbb{N}$. The relation between the auxiliary and physical fermions is given by
\begin{align}
  c_{k}^R\equiv
  \begin{cases}
    c_k \quad & k \geq 0
    \\
    d_k \quad & k<0        
  \end{cases}\quad,\quad
 c_{k}^L\equiv
  \begin{cases}
    c_k \quad & k < 0
    \\
    d_k\quad & k\geq0        
  \end{cases}\quad.
\end{align} 
The commutation relation thus reads $[b_q^i,{b_{q'}^{i'}}\dag]=\delta_{i,i'}\delta_{q,q'}$, with $i,i'=L,R$. 

The interaction term of the Hamiltonian given in Eq. (\ref{Liniarized_Hamiltonian}) can now be decoupled into left- and right-moving fluctuations if we assume the interaction to be long-ranged in position space, i.e. $v_q/v_F\ll1$ for $q\gg q_c$, where  $q_c\ll k_F$ denotes the interaction cutoff.\cite{Tom50} This suppresses scattering processes from one branch to the other, i.e. $g_1$-processes according to the $g$-ology model, due to $v_{\pm2k_F}\approx0$. A somehow counter-intuitive approach is to assume a delta-interaction in position space. A RG analysis then shows that repulsive $g_1$-processes scale to zero, i.e. are marginally irrelevant.\cite{Sol79}  The so-called $g_2$-processes, associated with the density-density interaction between left- and right-movers, are then renormalized to $g_2^*=g_2-g_1$.\cite{FootDe} 

To also express the kinetic term by the newly introduced bosonic operators, the Kronig relation has to be employed,
\begin{align}
\label{Kronig_Relation}
\sum_kk({c_k^R}\dag c_k^R-{c_k^L}\dag c_k^L)=\sum_{q>0}q({b_q^R}\dag b_q^R+
{b_q^L}\dag b_q^L)
\end{align}
where we neglected contributions of the  number-of-particle operator of the left- and right-movers.\cite{FootTL} The resulting TL model resembles the fixed point Hamiltonian of 1D interacting electron systems,\cite{Hal81}
\begin{align}
\label{Bosonized_Hamiltonian}
\begin{split}
H&=v_F\sum_{q>0}q(1+\frac{v_q}{2\pi v_F})({b_q^R}\dag b_q^R+
{b_q^L}\dag b_q^L)\\
&+\frac{1}{4\pi}\sum_{q>0}qv_q({b_q^R}b_q^L+
{b_q^L} b_q^R+{b_q^R}\dag {b_q^L}\dag+
{b_q^L}\dag {b_q^R}\dag).
\end{split}
\end{align}

Note that the kinetic term is renormalized by scattering processes which conserve the energy, denoted as $g_4$-processes in the $g$-ology model.\cite{Sol79} Also in the flow equation treatment of the $n$-orbital model, one has to include scattering terms which conserve the number of electrons above and below the Fermi points to the diagonal Hamiltonian.\cite{Weg94} 

\subsection{Solution via flow equation}

Since the Hamiltonian in Eq. (\ref{Bosonized_Hamiltonian}) is bilinear in bosonic operators and the left- and right-moving fluctuations are only coupled by the the same wave number, diagonalization is straightforward via Bogoljubov transformation. But we want to diagonalize the Hamiltonian via flow equations, i.e. we perform a continuous unitary transformation parametrized by the flow parameter $\ell$. 

In differential form, the infinitesimal transformations read $\partial_\ell H(\ell)=[\eta(\ell),H(\ell)]$, characterized by the anti-Hermitian generator $\eta(\ell)=-\eta\dag(\ell)$. The task is to properly define the generator such that the interaction term vanishes in the limit $\ell\to\infty$. Wegner proposed to choose $\eta=[H_0,H]$, where $H_0$ denotes the diagonal Hamiltonian which is assumed to be solved. This choice guarantees that the trace of the squares of the off-diagonal elements is a monotonically decreasing function of $\ell$, i.e. $\partial_\ell\text{Tr}[(H-H_0)^2]\leq0$. 

But  the above choice of $\eta$ also induces energy-scale separation, i.e. for a simple model Wegner showed that the energy difference between the eigenstates of $H_0$, which are just being decoupled, is given by $\ell^{-1/2}$. Even though all generators which diagonalize the system for $\ell\to\infty$ are essentially equivalent, a systematic decoupling scheme becomes important when approximations are involved. For a recent review on the flow equation method, see Ref. \onlinecite{Weg01}.

In order to abbreviate the notation we define $b_q\equiv b_q^R$ and $b_{-q}\equiv b_q^L$ for $q>0$. Neglecting the shift of the ground-state energy, the $\ell$-dependent Hamiltonian then takes the following form:
\begin{align}
\notag
H(\ell)&=\sum_{q\neq0}\omega_q(\ell)b_q\dag b_q+\frac{1}{2}\sum_{q\neq0}u_q(\ell)(b_qb_{-q}+b_{-q}\dag b_q\dag)\\
\label{FlowHamiltonian_TL}
&\equiv H_0+V
\end{align}
The initial conditions are given by $\omega_q^0\equiv\omega_q(\ell=0)=v_F|q|(1+v_q/2\pi v_F)$ and $u_q^0\equiv u_q(\ell=0)=|q|v_q/2\pi$. The generator of the infinitesimal transformations is canonically defined as
\begin{align}
\eta=[H_0,H]=-\sum_{q\neq0}\omega_{q}u_q(b_qb_{-q}-b_{-q}\dag b_q\dag)\quad,
\end{align}
where we already used the fact that the flow equations preserve the symmetry between left- and right-movers, i.e. $\omega_q(\ell)=\omega_{-q}(\ell)$ and $u_q(\ell)=u_{-q}(\ell)$ for all $\ell$. The above choice is likely to eliminate the interaction term $V$ for $\ell\to\infty$,\cite{Weg94} but also different generators are possible.\cite{Ste98}

The commutator $[\eta,H]$ yields the following contributions:
\begin{align}
[\eta,H_0]&=-\sum_{q\neq0}2\omega_{q}^2u_q(b_qb_{-q}+b_{-q}\dag b_q\dag)\\
[\eta,V]&=-\sum_{q\neq0}2\omega_{q}u_q^2(b_qb_q\dag+b_q\dag b_q)
\end{align}
The flow equations $\partial_\ell H=[\eta,H]$ then read
\begin{align}
\label{FlowEquations_TL}
\partial_\ell\omega_q=-4\omega_qu_q^2\quad,\quad\partial_\ell u_q=-4\omega_q^2u_q\quad.
\end{align}
Obviously $\omega_q^2-u_q^2=\text{const}$ and with $u_q(\ell=\infty)=0$ we have $\tilde{\omega}_q\equiv\omega_q(\ell=\infty)=(\omega_q^2(\ell)-u_q^2(\ell))^{1/2}$. Inserting the initial conditions for $\ell=0$ yields the well-known result $\tilde{\omega}_q=v_F|q|\sqrt{1+v_q/\pi v_F}$.\\ 

In order to investigate the flow of observables we need to know the $\ell$-dependence of $\eta_q\equiv-\omega_qu_q$. From the above equations we obtain 
\begin{align}
\notag
\eta_q(\ell)=-\frac{\text{sgn}(v_q)\tilde{\omega}_q^2}{2\sinh(4\tilde{\omega}_q^2\ell+C_q)}\quad,\quad\sinh(C_q)=\frac{\tilde{\omega}_q^2}{2\omega_q^0|u_q^0|}\quad.
\end{align}
The explicit $\ell$-dependence of the parameters $\omega_q$ and $u_q$ is given by
\begin{align}
\label{SolutionwqTL}
\omega_q(\ell)&=\cosh(4E_q(\ell))\omega_q^0+\sinh(4E_q(\ell))u_q^0\\\label{SolutionvqTL}
u_q(\ell)&=\cosh(4E_q(\ell))u_q^0+\sinh(4E_q(\ell))\omega_q^0
\end{align}
where we defined $E_q(\ell)\equiv\int_0^\ell d\ell'\eta_q(\ell')$. One can now convince oneself that these solutions indeed yield the correct boundary values for $\ell\rightarrow\infty$.

Since $[\eta(\ell),\eta(\ell')]=0$ we can also calculate the unitary operator $U$ that diagonalizes the TL Hamiltonian. Generally it is given by $U\equiv U(\ell=\infty)$ with $U(\ell)=\mathcal{L}\exp(\int_0^{\ell}d\ell'\eta(\ell'))$ where the operator $\mathcal{L}$ denotes the $\ell$-ordering operator, defined in the same way as the time-ordering operator. We obtain
$U=\exp(\sum_{q\neq0}E_q^* (b_qb_{-q}-b_{-q}\dag b_{q}\dag))$
with the familiar relation $\tanh(4E_q^* )=-v_q/(2\pi v_F+v_q)$, where we defined $E_q^* \equiv E_q(\ell=\infty)$.\\

To determine the flow of the observables we will use the representation of the 1D fermionic field operator of the left- and right-movers, $\psi_{L/R}(x)$, which involves the operators of the bosonic density fluctuation.\cite{Lut74,Hal81} It is given by $\psi_{L/R}(x)=F_{L/R}e^{\phi_{L/R}(x)}$, where the operator $F_{L/R}$ lowers the number of left- and right-movers by one, respectively and commutes with the bosonic operators ${b_q}^{(\dagger)}$, i.e. $[b_q^{(\dagger)},F_{L/R}]=0$.\cite{FootKl} The phase field is defined as 
\begin{align}
\phi_{L/R}(x)\equiv\sum_{q>0}n_q^{-1/2}(b_{\mp q}e^{\mp iqx}-b_{\mp q}^\dagger e^{\pm iqx})\quad,
\end{align}
where, for convenience, we omitted the ultraviolet convergence factor. Due to the preserved symmetry between left- and right-movers, only the transformation of the right-moving field needs to be considered. 

Since $U(\ell)$ only consists of bosonic operators and thus commutes with $F_R$, we have $U\psi_R(x)U^{\dagger}=F_Re^{U\phi_{R}(x)U\dag}$ for all $\ell$. Therefore it suffices to consider the flow equations for the phase field, i.e. $\partial_\ell\phi_R(x)=[\eta,\phi_R(x)]$. During the flow of $\phi_R(x,\ell)\equiv\sum_{q>0}\tilde\phi_q^R(x,\ell)$ different wave numbers do not mix and we are allowed to limit ourself to the flow of $\tilde\phi_q^R(x,\ell)\equiv\varphi_q^R(\ell)\phi_q^R(x)+\varphi_q^L(\ell)\phi_q^L(x)$ with 
\begin{align}
\label{phi_Q}
\phi_q^{L/R}(x)\equiv \big(b_{\mp q}e^{\mp iqx}-b_{\mp q}^\dagger e^{\pm iqx}\big)\quad. 
\end{align}
This yields the following flow equations:
\begin{align}
\partial_\ell\varphi_q^R=-2\eta_q \varphi_q^L\quad,\quad\partial_\ell \varphi_q^L=-2\eta_q\varphi_q^R
\end{align}
With the initial conditions $\varphi_q^R(\ell=0)=n_q^{-1/2}$ and $\varphi_q^L(\ell=0)=0$ the solution is given by
\begin{align}
\varphi_q^R(\ell)&=n_q^{-1/2}\cosh(2E_q(\ell))\quad,\notag\\
\varphi_q^L(\ell)&=-n_q^{-1/2}\sinh(2E_q(\ell))\quad.\notag
\end{align}
In the limit $\ell\to\infty$ we recover the well-known result for the transformed field operator
\begin{align}
\psi_R(x)&\to F_R\exp\Big(\sum_{q>0}\big(\frac{c_q}{\sqrt{n_q}}(b_qe^{iqx}-b_{q}\dag e^{-iqx})\notag\\
&-\frac{s_q}{\sqrt{n_q}}(b_{-q}e^{-iqx}-b_{-q}\dag e^{iqx})\big)\Big)\quad,
\label{FermionField}
\end{align}
with $c_q\equiv\cosh(2E_q^*)$, $s_q\equiv\sinh(2E_q^* )$, and  
$s_q^2=(\omega_q^0/\tilde{\omega}_q-1)/2$, see e.g. Ref. \onlinecite{Sch97}. Correlation functions in position-time space are now easily calculated.\\

\subsection{Approximations}

It is not surprising that the flow equation approach could be applied in a rather straightforward manner because we took advantage of the bosonization technique. Now we want to calculate the commutator $[\eta,\psi_R(x)]$ directly, employing the usual representation $\psi_R(x)=L^{-1/2}\sum_{k}e^{ikx}c_k^R$. With the definition of Eq. (\ref{phi_Q}) this yields
\begin{align}
\notag
[\eta,\psi_R(x)]=-2\sum_{q>0}\eta_{q}n_q^{-1/2}\phi_q^L(x)\psi_R(x)\quad.
\end{align}
Further we have $[\eta,\phi_q^{L/R}(x)]=-2\eta_q\phi_q^{R/L}(x)$. This shows that the flow equations $\dl \psi_R(x)=[\eta,\psi_R(x)]$ do not close but generate an infinite series of operators. We thus make the following ansatz for the fermionic field:
\begin{align}
\notag
\psi_R(x,\ell)&=\psi_R(x)\Big[g(\ell)+\sum_{q>0}\big(\varphi_q^R(\ell)\phi_q^R(x)\\
&+\varphi_q^L(\ell)\phi_q^L(x)\big)\Big]\label{TruncField}
\end{align} 
The initial conditions are given by $g(\ell=0)=1$ and $\varphi_q^{L/R}(\ell=0)=0$. The above ansatz implies that we only keep track of the first non-trivial term generated by the flow equations.

The flow equations depend on the decoupling scheme of the higher order terms. If one neglects bilinear terms in the bosonic operators without normal ordering them, the parameter $g$ is left un-renormalized, i.e. $g=1$ for all $\ell$. The flow equations then read
\begin{align}
\notag
\partial_\ell\varphi_q^R&=-2\eta_q \varphi_q^L\quad,\\
\partial_\ell \varphi_q^L&=-2\eta_q\varphi_q^R-2n_q^{-1/2}\eta_q\quad.
\label{DiffField}
\end{align} 
An analytic solution is now possible.
In the limit $\ell\to\infty$ the final result is given by
\begin{align}
\notag
\varphi_q^R(\ell=\infty)=n_q^{-1/2}\tilde{c}_q\quad,\quad
\varphi_q^L(\ell=\infty)=-n_q^{-1/2}s_q
\quad,
\end{align}
where we abbreviated $\tilde{c}_q\equiv c_q-1$.

Expanding also the fermionic field up to linear bosonic operators, i.e. $\psi_R(x)\to F_R(1+\sum_{q>0}n_q^{-1/2}\phi_q^R(x))$, we see that the flow equation approach yields the correct result of Eq. (\ref{FermionField}) up to terms which are linear in the bosonic operators, i.e. $U\psi_R(x)U\dag\to F_R(1+\sum_{q>0}n_q^{-1/2}(c_q\phi_q^R(x)-s_q\phi_q^L(x))$. Note that the Bogoljubov-coefficients $s_q$ and $c_q$ are exactly recovered.

Consider now the transformation of the fermionic ladder operators of the right-movers $c_k^R$. The flow equations $\dl c_k^R=[\eta,c_k^R]$ do not close and we will again truncate the series after the terms which are linear in the bosonic operators,
\begin{align}
c_k^R(\ell)=g(\ell)c_k^R+\sum_{q>0}\big(\varphi_q^R(\ell)\phi_{q,k}^R+\varphi_q^L(\ell)\phi_{q,k}^L\big)\quad,
\end{align}
with the initial conditions $g(\ell=0)=1$ and $\varphi_q^{L/R}(\ell=0)=0$ and where we defined $\phi_{q,k}^{L/R}\equiv(b_{\mp q}c_{k\pm q}^R-b_{\mp q}\dag c_{k\mp q}^R)$.

Neglecting terms which are bilinear in the bosonic operators in their non-normal ordered form leads to the same differential equations as for the flow parameters of the truncated fermionic field operator. The solution is thus again given by $g=1$ for all $\ell$ and 
\begin{align}
\notag
\varphi_q^R(\ell=\infty)=n_q^{-1/2}\tilde{c}_q\quad,\quad
\varphi_q^L(\ell=\infty)=-n_q^{-1/2}s_q
\quad.
\end{align}

So far we do not know what effect the above truncation scheme has got on physical quantities. For this reason we will now calculate the occupation function in momentum space within this approximation, i.e. $n_k^{R,1}\equiv\langle FDS|{c_k^R}\dag(\ell=\infty) c_k^R(\ell=\infty)|FDS\rangle$, where $|FDS\rangle$ denotes the ground-state of the Fermi-Dirac sea. We obtain the following result:
\begin{align}
\label{Occupation_1_TL}
n_k^{R,1}&=(1-\sum_{q>0}\frac{\tilde{c}_q}{n_q})^2\Theta(k_F-k)\notag\\
&+2(1-\sum_{q>0}\frac{\tilde{c}_q}{n_q})\sum_{q>0}\frac{\tilde{c}_q}{n_q}\Theta(k_F-q-k)\notag\\
&+\sum_{q>0}\frac{\tilde{c}_q}{n_q}\sum_{q'>0}\frac{\tilde{c}_{q'}}{n_{q'}}\Theta(k_F-q-q'-k)
\notag\\
&+\sum_{q>0}\frac{\tilde{c}_q^2}{n_q}\Theta(k_F-q-k)+\sum_{q>0}\frac{s_q^2}{n_q}\Theta(k_F+q-k)\quad,
\end{align}
with the step function $\Theta(k)$ which equals one for $k\geq0$ and zero otherwise.

Considering $k=k_F+\tilde{q}$ with $\tilde{q}>0$, only the last term of the above equation contributes to $n_k^{R,1}$. It is now convenient to work with a finite interaction cutoff $q_c\equiv2\pi n_c/L$ and choose the interaction potential as a step function in momentum space, i.e. $s\equiv s_q$ for $q\leq q_c$ and zero otherwise. Like this, we avoid ultraviolet divergences and are able to obtain analytic results. With $\sum_{n=1}^{n_c}1/n\to \ln{n_c}+C$ for $n_c\to\infty$, where $C$ is Euler's constant, we find in the thermodynamic limit $L\to\infty$, $N/L=\text{const}$ the following expression:
\begin{align}
\label{BZleft}
n_{k_F+\tilde{q}}^{R,1}&=\sum_{q>0}\frac{s_q^2}{n_q}\Theta(q-\widetilde{q})\to-s^2\ln(\tilde{q}/q_c)\notag\\
&\to(1-(\tilde{q}/q_c)^{2s^2})/2 
\end{align}   
The last limit was taken in the case of weak coupling, where we assumed that the logarithm stems from a power-law expansion. In this perturbative regime we thus recover the well-known algebraic behavior around the Fermi point $k_F$. We also find the remarkable result that the above truncation scheme yields the {\em exact} anomalous dimension $\alpha\equiv2s^2$.

For wave numbers below the Fermi point, i.e. for $k=k_F-\tilde{q}$ with $\tilde{q}>0$, the calculations are not so plane and we will have to make further approximations. In the following, we have to discard $\sum_{q,q'>0}\tilde{c}_q\tilde{c}_{q'}/(n_qn_{q'})(\Theta(k_F+q-k)\Theta(k_F+q'-k)+\Theta(k_F-q-q'-k)-\Theta(k_F-q-k)\Theta(k_F-q'-k))$ and the constant term $\sum_{q>0}(\tilde{c}_q^2+s_q^2)/n_q$. The neglect of the first terms can be justified since they resemble higher order contributions to the logarithmic expansion and by only considering the first non-trivial term in the expansion of $\psi_R(x)$, these contributions cannot be accounted for. The second, constant term implies that the neglected operators should be included in the flow more adequately. Normal ordering before neglecting these operators would result in a renormalization of $g$ which would also give rise to a constant contribution, partially canceling the first one. Nevertheless, an analytic treatment is then not possible anymore and we do not want to extend the present approach.  For a detailed discussion on the observable flow in the case of a numerically solvable, but non-trivial model, see Ref. \onlinecite{Sta02}.

With these approximations, one then obtains for the whole regime in the limit of small coupling
\begin{align}
\label{nk}
n_{k_F+\tilde{q}}^{R,1}=1/2-\text{sgn}(\tilde{q})(|\tilde{q}|/q_c)^{\alpha}/2\quad.
\end{align}

The pre-factor $1/2$ associated with the power-law behavior is also recovered from simple perturbation theory.\cite{Sch97} This factor could not be recovered unambiguously from the calculations of the approximate occupation number of the $n$-orbital model.\cite{Weg94}\\

Finally, we want to check if also dynamical quantities can be recovered from the above truncation scheme. For this we will calculate the approximate Green function defined as
\begin{align}
iG_R^{<,1}(x,t)&\equiv\langle FDS|{\psi_R}\dag(x=0,\ell=\infty)\notag\\
\label{Green_Trunk_TL}
&\times e^{iH^* t}\psi_R(x,\ell=\infty)e^{-iH^* t}|FDS\rangle\quad,
\end{align}
with $H^*\equiv H(\ell=\infty)$.

In order to evaluate the time dependence of the fermionic field of the right-movers we have to work with a constant potential in momentum space, i.e. $v\equiv v_q$ for all $q$. This yields a linear energy dispersion for the fixed point Hamiltonian, i.e. $H^*=v_c\sum_{q\neq0}|q|b_q\dag b_q$ with the renormalized Fermi or charge velocity $v_c\equiv v_F(1+v/(\pi v_F))^{1/2}$. With the help of the Kronig relation of Eq. (\ref{Kronig_Relation}), the time dependence of the fermionic and bosonic ladder operators in the Heisenberg picture at $\ell=\infty$ is given by $c_k^R(t)=c_k^Re^{-ikv_ct}$ and $b_q(t)=b_qe^{-i|q|v_ct}$.
 
The Green function of Eq. (\ref{Green_Trunk_TL}) can now be expressed as a function of the conformal variables $\xi^{L/R}\equiv x\pm v_ct$ and yields
\begin{align}
\label{GreensfunctionTL}
\notag
iG_R^{<,1}(\xi^R,\xi^L)&=iG_R^{<,0}(\xi^R)\Big[\big(1-\sum_{q>0}\frac{\tilde{c}_q}{n_q}(1-e^{-iq\xi^R})\big)^2\\
&+\sum_{q>0}\frac{\tilde{c}_q^2}{n_q}e^{-iq\xi^R}+\sum_{q>0}\frac{s_q^2}{n_q}e^{iq\xi^L}\Big]\quad,
\end{align}
with $iG_R^{<,0}(\xi^R)\equiv L^{-1}\sum_ke^{ik\xi^R}\Theta(k_F-k)$.

An important consistency check of the above truncation schemes is given by the fact that $n_k^{R,1}=\int_{-L/2}^{L/2}dxe^{-ikx}iG_R^{<,1}(x,t=0)$. But 
working with a constant potential in momentum space, i.e. a delta-potential in position space,  yields the well-known ultraviolet divergences. In the following we will therefore introduce an ultraviolet cutoff and also label it with $q_c$.\\ 

A physical observable is given by the occupied density of states $\rho_R^<(\omega)$ which is observed in photo-emission experiments. In the above approximation, it is given by
\begin{align}
\rho_R^{<,1}(\omega)\equiv\frac{1}{2\pi}\int dte^{i\omega t}iG_R^{<,1}(x=0,t)\quad.
\end{align}
Neglecting the same terms as we did in the case of the calculation of the occupation number, we obtain with $\omega\equiv v_ck_F-\tilde{\omega}$ in the thermodynamic limit
\begin{align}
\rho_R^{<,1}(\omega)=\Theta(\tilde{\omega})\big[1+2s^2\ln(\tilde{\omega}/(v_cq_c)\big]/(2\pi v_c)\quad.\label{BulkDOS}
\end{align} 
In the limit of small coupling we thus obtain the well-known algebraic suppression at the renormalized Fermi energy, again governed by the {\em exact} anomalous dimension $\alpha=2s^2$.\cite{FootEx} Note that for the static correlation function $n_k^{R,1}$ the anomalous dimension is being accounted for by either left-movers ($k>k_F$) or right-movers ($k<k_F$) whereas both branches contribute in equal parts in case of the dynamic correlation function $\rho_R^{<,1}(\omega)$. Finally, we want to mention that Eqs. (\ref{BZleft}) and (\ref{BulkDOS}) are also obtained within RPA perturbation theory.\cite{Med96}

\section{Tomonaga-Luttinger Model with Impurity}

The two examples of the previous section showed that the flow equation approach can yield the exact anomalous dimension even within a rather crude truncation scheme. We will now add an impurity to the system and employ the same truncation scheme as above, i.e. we use the fermionic representation of the field operators to represent the impurity and only keep track of the first non-trivial term. 

\subsection{The model}
Departing from the Hamiltonian of Eq. (\ref{TL-Hamiltonian}), the one-body potential shall now be given by $U(x)=\lambda_0\delta(x)$. Performing the same approximations that led to the TL model and employing the same notation as in the previous section, we arrive to the following Hamiltonian: 
\begin{align}
\label{FlowHamiltonian_TLwithImpurity}
H&=\sum_{q\neq0}\omega_q^0b_q\dag b_q+\frac{1}{2}\sum_{q\neq0}u_q^0(b_qb_{-q}+b_{-q}\dag b_q\dag)+\lambda_0\psi\dag\psi\notag\\
&\equiv H_0+H_{ee}+H_i\quad,
\end{align}
with $\psi\equiv\psi(x=0)$. The impurity term $H_i$ is usually split up into two contributions, i.e. a forward scattering and a backward scattering part. With $\psi\equiv\psi_L+\psi_R$ they read
\begin{align}
\label{ForwardBackward_Impurity}
\begin{split}
H_i&=\lambda^F(\psi_R\dag\psi_R+\psi_L\dag\psi_L)+\lambda^B(\psi_R\dag\psi_L+\psi_L\dag\psi_R)\notag\\
&\equiv H_F+H_B\quad.
\end{split}
\end{align}
The bosonic operators $b_q^{(\dagger)}$ and the fermionic field operators are thus correlated via the following commutation relation:
\begin{align}
\notag
[b_q,\psi_{L/R}]=[b_q^{\dagger},\psi_{L/R}]=-\Theta(\mp q)n_q^{-1/2}\psi_{L/R}
\end{align}

The forward scattering contribution $H_F$ can be expressed by the bosonic density fluctuations, i.e. $H^F=L^{-1}\sum_{q\neq0}n_q^{1/2}\lambda_q^F(b_q+b_q\dag)$, where we neglected the term proportional to the number-of-particle operator of the left- and right-movers. The initial conditions are given by $\lambda_q^F(\ell=0)=\lambda_0$ and $\lambda^B(\ell=0)=\lambda_0$.
 
\subsection{Solution via flow equations}
We will now set up flow equations which will preserve the form of the Hamiltonian of Eq. (\ref{FlowHamiltonian_TLwithImpurity}). This strategy has already proved to be successful in the context of the spin-boson model.\cite{Keh96}
The generator of the infinitesimal transformations shall consist of three parts $\eta\equiv\eta^{ee}+\eta^F+\eta^B$ according to the canonical generator $\eta_c\equiv[H_0,H]$, but with generalized parameters which will be determined later,
\begin{align}
\eta^{ee}&=\sum_{q\neq0}\eta_q^{ee}(b_qb_{-q}-b_{-q}\dag b_q\dag)\quad,\quad
\eta^F=\sum_{q\neq0}\eta_q^F(b_q-b_q\dag)\notag\\
\eta^B&=\sum_{q\neq0}\eta_q^B\big((b_q\dag\psi_R\dag\psi_L+
\psi_R\dag\psi_Lb_q)-h.c.\big)\quad.\notag
\end{align}
Notice that $\eta^B$ is not normal ordered in the usual sense such that the fermionic creation operators are left from the fermionic annihilation operators. For the following, we thus want to define normal ordering such that the creation (annihilation) operators $b_q^{\dagger}$ $(b_q)$ are left (right) from the $2k_F$-operators $\psi_R\dag\psi_L$ and $\psi_L\dag\psi_R$, respectively. By this, we take on the standpoint that the bosonic operators $b_q^{\dagger}$ are independent entities with no internal structure.

To clarify the approach we will now present the several contributions of the commutator $[\eta,H]$.
The commutator $[\eta^{ee},H]$ yields the following contributions:
\begin{align}
[\eta^{ee},H_0]&=2\sum_{q\neq0}\omega_{q}\eta_{q}^{ee}(b_qb_{-q}+b_{-q}\dag b_q\dag)\\
[\eta^{ee},H_{ee}]&=2\sum_{q\neq0}\eta_{q}^{ee}u_q(b_qb_q\dag+b_q\dag b_q)\\
[\eta^{ee},H_F]&=2L^{-1}\sum_{q\neq0}n_q^{1/2}\eta_q^{ee}\lambda_{-q}^F(b_q+b_q\dag)\\
[\eta^{ee},H_B]&=2\sum_{q\neq0}\eta_q^{ee}\lambda^B\text{sgn}(q)n_q^{-1/2}(\Psi_q+h.c.)\label{etaeeHb}\\\notag
&-2\sum_{q\neq0}\eta_q^{ee}\lambda^Bn_q^{-1}(\psi_R\dag\psi_L+h.c.)
\end{align}
In Eq. (\ref{etaeeHb}) we defined the operators $\Psi_q\equiv (b_q\dag\psi_R\dag\psi_L-\psi_R\dag\psi_Lb_q)$, whose contributions to the flow will be suppressed by a proper choice of $\eta_q^B$.
The relevant contributions of $[\eta^F,H]$ are given by
\begin{align}
[\eta^F,H_0]&=\sum_{q\neq0}\eta_q^F\omega_q(b_q+b_q\dag)\quad,\\
[\eta^F,H_{ee}]&=\sum_{q\neq0}\eta_q^Fu_q(b_q+b_q\dag)\quad.
\end{align}
The commutator $[\eta^B,H]$ yields the following contributions ($[\eta^B,H_{F}]=0$):
\begin{align}
\label{etaBH0_Imp}
[\eta^B,H_0]&\approx-\sum_{q\neq0}\eta_q^B\omega_q(\Psi_q+h.c.)\\
[\eta^B,H_{ee}]&\approx\sum_{q\neq0}(\eta_{-q}^Bu_q-\eta_q^B\kappa)(\Psi_q+h.c.)\label{etaBHee_Imp}\\\notag
	&+2\sum_{q\neq0}\eta_q^Bu_q\text{sgn}(q)n_{q}^{-1/2}(\psi_R\dag\psi_L+h.c.)\\
[\eta^B,H_B]&=4\sum_{q\neq0}\eta_{q}^B\lambda^B\text{sgn}(q)n_{q}^{-1/2}\psi_R\dag\psi_R\psi_L\dag\psi_L\label{etaBHB_Imp}\\\notag
&+2\rho\sum_{q\neq0}\eta_{q}^B\lambda^B\text{sgn}(q)n_{q}^{-1/2}(\psi_R\dag\psi_R+\psi_L\dag\psi_L)\\\notag
&+2\rho\sum_{q\neq0}\eta_{q}^B\lambda^B\big(b_q\dag(\psi_R\dag\psi_R-\psi_L\dag\psi_L)+h.c\big)
\end{align}

In Eq. \ref{etaBHee_Imp}, we introduced the $\ell$-dependent parameter $\kappa\equiv\sum_{q\neq0}n_{q}^{-1}u_q$.
The symbol $\approx$ in Eqs. (\ref{etaBH0_Imp}) and (\ref{etaBHee_Imp}) signifies that we neglect normal ordered operators which consist of  two bosonic operators and one $2k_F$-operator ($\psi_R\dag\psi_L$ or $\psi_L\dag\psi_R$). This is our truncation scheme which should be valid for small electron-electron interaction $v_q$ as was pointed out in the previous section.

We will now look more closely at Eq. (\ref{etaBHB_Imp}). First, we note that the factor $\rho\equiv N/L$ emerges from the regularization of the anti-commutation relation $\{\psi_{L/R},\psi_{L/R}\dag\}=L^{-1}\sum_k\approx N/L$, having in mind the restriction of the sum to the first Brillouin zone. Secondly, we see that new interaction terms are generated which can be expressed by the bosonic density fluctuations of the left- and right-movers, i.e. $H_R\equiv L^{-1}\sum_{q,q'\neq0}(R_{q,q'}^1b_{q}\dag b_{q'}+R_{q,q'}^2(b_{q}b_{q'}+b_{q'}\dag b_{q}\dag))$ with general coupling matrices $R_{q,q'}^{1/2}$. This contribution had to be included into the flow of the Hamiltonian. However, there is evidence based on investigations of the TL model with open boundaries that these terms represent boundary effects which - in the thermodynamic limit - do not alter the universal features of the model.\cite{Med00} The evolution of $H_R$ shall thus be neglected.

We will now specify the flow. The generators $\eta^{ee}$ and $\eta^F$ are determined according to the canonical generator $\eta_c\equiv[H_0,H]$, i.e. $\eta_q^{ee}=-\omega_qu_q$ and $\eta_q^F=-\omega_qn_q^{1/2}L^{-1}\lambda_q^F$. The generator $\eta^B$ is chosen in order to eliminate the newly generated terms that contain the operator $(\Psi_q+h.c.)$. This yields
\begin{align}
\label{etaB}
\eta_q^B=\text{sgn}(q)n_q^{-1/2}\lambda^B\frac{2\eta_q^{ee}}{\omega_q+u_q+\kappa}\quad.
\end{align}
The constant $\kappa$, which stems from normal ordering, acts as a low-momentum cutoff and can be neglected in the asymptotic regime since $\kappa\to\ell^{-1/2}$. For $\ell=0$, we obtain $\kappa\propto vq_c$ where $v$ is a measure of the electron-electron interaction and $q_c$ denotes the interaction cutoff. For attractive interaction, $v<0$, and large $|v|$ and $q_c$, this might lead to a singular expression of Eq. (\ref{etaB}). For the following discussion, this parameter regime has to be excluded. 

With $\partial_\ell H=[\eta,H]$, the flow equations for the bulk parameters read
\begin{align}
\label{FlowEquations_TLBulk}
\partial_\ell\omega_q&=-4\omega_qu_q^2\quad,\quad\partial_\ell u_q=-4\omega_q^2u_q\quad.
\end{align}
These are the same flow equations as in the case without the impurity. The solution is thus given by Eqs. (\ref{SolutionwqTL}) and (\ref{SolutionvqTL}). The asymptotic behavior of the coupling $u_q(\ell)$ turns out to be crucial for the following analysis. It can be deduced from the exact solution and yields $u_q(\ell)\to u_q^0(\tilde{\omega}_q/\omega_q^0)e^{-4\tilde{\omega}_q^2\ell}$ for $\ell\to\infty$.

The flow equations for the scattering terms read
\begin{align}
\dl \lambda^B&=\lambda^B\sum_{q>0}\Big(\frac{4\omega_qu_q}{n_q}-\frac{8\omega_qu_q^2}{n_q}\frac{1}{\omega_q+u_q+\kappa}\Big)\label{FlowEquations_TLBackImpurity}\quad,\\
\dl \lambda_q^F&=-\lambda_q^F(\omega_q^2+3\omega_q u_q)\notag\\
	&+2\rho(\lambda^B)^2\sum_{q'>0}n_{q'}^{-1}\frac{4\omega_{q'}u_{q'}}{\omega_{q'}+u_{q'}+\kappa}\label{FlowEquations_TLForwardImpurity_Imp}\quad.
\end{align}
Setting $\kappa=0$, we observe after some algebra that the flow of the impurity strength delicately depends on the sign of the interaction potential $v_q$. For attractive electron-electron interaction $v_q<0$ the absolute value of the backward scattering potential $\lambda^B$ decreases whereas it increases for repulsive interaction. We will now show that $|\lambda^B|$ tends to either zero or infinity which resembles the main result of Kane and Fisher.

\begin{figure}[t]
  \begin{center}
    \epsfig{file=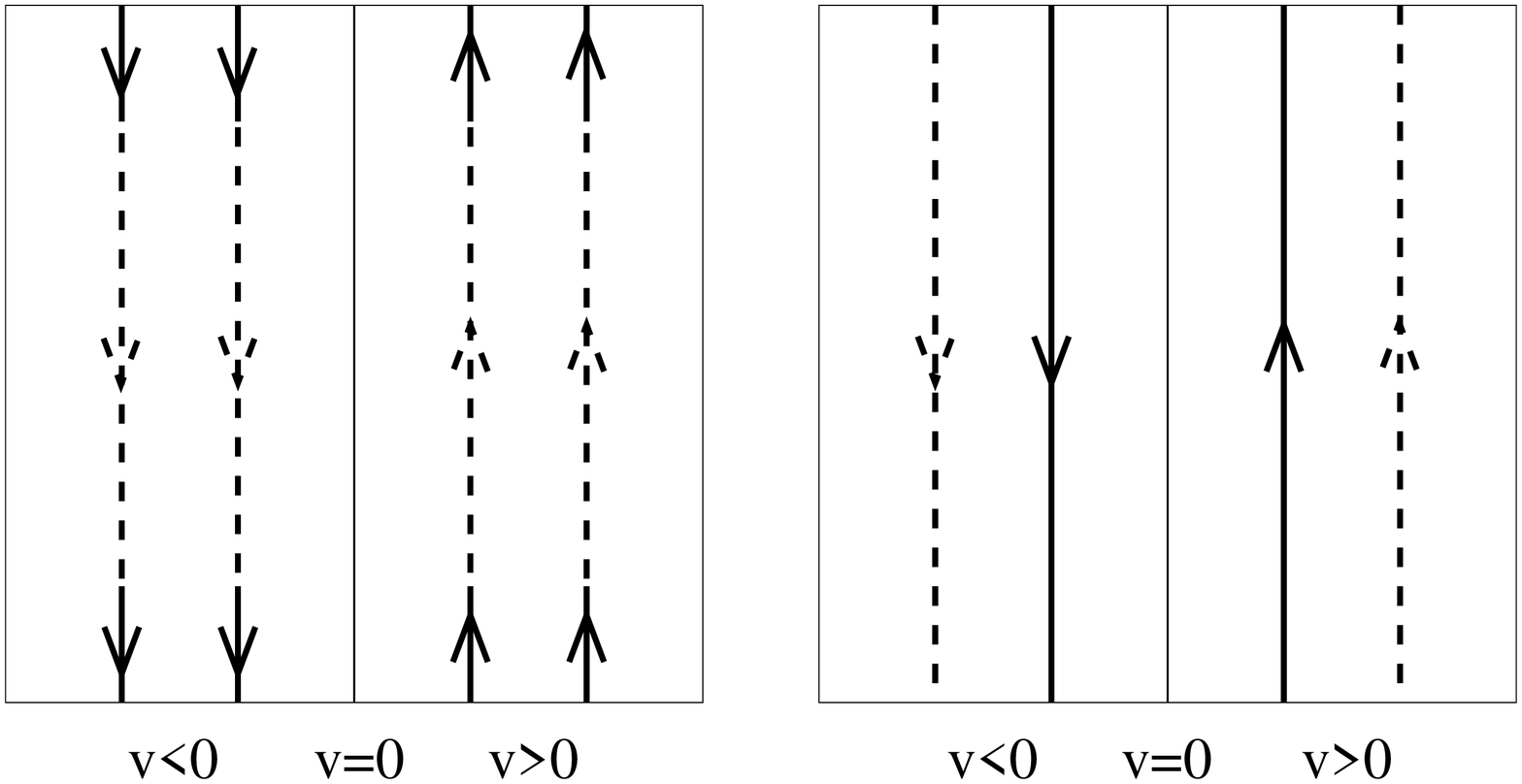,height=4cm,angle=0}
    \caption{The qualitative flow of $|\lambda^B|$ depending on the sign of the two-body interaction $v$ as it follows from perturbative RG analysis (l.h.s.) and from the flow equation approach (r.h.s.). The solid lines resemble analytic results whereas the dashed lines are inferred.}
    \label{FlowLb}
\end{center}
\end{figure}

Eq. (\ref{FlowEquations_TLBackImpurity}) is easily integrated and yields $\lambda^B(\ell)=\lambda^B(\ell=0)\exp(\int_0^{\ell}d\ell'S(\ell'))$, where we defined 
\begin{align}
\notag
S&\equiv\sum_{q>0}\Big(\frac{4\omega_qu_q}{n_q}-\frac{8\omega_qu_q^2}{n_q}\frac{1}{\omega_q+u_q+\kappa}\Big)\\
&\to4\sum_{q>0}\frac{u_q^0}{n_q}\frac{\tilde{\omega}_q^2}{\omega_q^0}e^{-4\tilde{\omega}_q^2\ell}
\end{align}
and the limit was taken for $\ell\to\infty$.
If we choose the electron-electron interaction potential as a step function with momentum cutoff $q_c$, i.e. $v_q=v\Theta(q_c-|q|)$,\cite{FootPot} then the sum can be performed in the thermodynamic limit,
\begin{align}
\notag
S\to4\tilde{\omega}^2\frac{u}{\omega}\int_0^{q_c} dqqe^{-4\tilde{\omega}_q^2\ell}
=\frac{u}{2\omega}\ell^{-1}(1-e^{-4\tilde{\omega}_{q_c}^2\ell}),
\end{align}
where we defined $\tilde{\omega}\equiv\lim_{q\to0}\tilde{\omega}_q/q$, $u\equiv\lim_{q\to0}u_q^0/q$ and $\omega\equiv\lim_{q\to0}\omega_q^0/q$. The asymptotic behavior of the backward scattering impurity potential is thus given by 
\begin{align}
\lambda^B\to\lambda_*^B\ell^{\alpha_B^*/2}
\end{align}
where $\alpha_B^*\equiv v/(2\pi v_F+v)$ and $\lambda_*^B$ denotes the $\ell$-independent asymptotic part of $\lambda^B$. 
This leads to the scenario described above and we want to stress that the asymptotic behavior is independent of the initial condition, i.e. holds for arbitrary initial impurity strength $\lambda_0$. In Fig. \ref{FlowLb}, the qualitative flow of $|\lambda^B|$ depending on the sign of the two-body interaction $v$ is shown as it follows from the one-loop perturbative RG analysis of Kane and Fisher (l.h.s.) and from the flow equation approach (r.h.s.). The solid lines resemble analytic results whereas the dashed lines are inferred.

Note that the phase separation is due to the asymptotic flow in which the system is already almost decoupled, i.e. the two-body interaction $u_q(\ell)$ is exponentially small. The localization phenomena is thus a genuine low-energy phenomena. Under these considerations it appears plausible that for real (finite) samples this asymptotic regime may be hard to reach.\cite{Med02} 

What is still left to verify is that the forward scattering potential $\lambda_q^F$ will be negligible compared to $\lambda_q^B$ for $\ell\to\infty$. The asymptotic behavior of the inhomogeneous part of Eq. (\ref{FlowEquations_TLForwardImpurity_Imp}) is given by 
\begin{align}
\sum_{q>0}n_{q}^{-1}\frac{4\omega_{q}u_{q}}{\omega_{q}+u_{q}+\kappa}&\to4\tilde{\omega}\frac{u}{\omega}\int_0^{q_c}dqe^{-4\tilde{\omega}_q^2\ell}\notag\\
&=\pi^{1/2}\frac{u}{\omega}\ell^{-1/2}\text{erf}(2\tilde{\omega}_{q_c}\sqrt{\ell})\quad,
\end{align}
with the error function $\text{erf}(x)\equiv2\pi^{-1/2}\int_0^x dx'e^{-{x'}^2}\to1-e^{-x^2}/(x\pi)$ for $x\to\infty$.
The differential equation of Eq. (\ref{FlowEquations_TLForwardImpurity_Imp}) can thus be solved in leading order as $\ell\to\infty$ to yield $\lambda_q^F=\pi^{1/2}\rho(u/\omega)(\lambda_*^B/\tilde{\omega}_q)^2\ell^{\alpha_B^*-1/2}$.

Since $\alpha_B^*=v/(2\pi v_F+v)$, the forward scattering impurity strength tends to zero as $\ell\to\infty$ for $v\leq2\pi v_F$. For $v>2\pi v_F$ - one must question whether the truncation scheme still holds for this coupling regime - $\lambda_q^F$ goes to infinity, but with a weaker power-law behavior than the backward scattering potential. The forward scattering term will therefore always resemble the less relevant contribution to the asymptotic flow.

Analyzing the above equation for finite system size $L$, we find that the maximum value of $|\lambda^F|$ increases exponentially as a function of the initial value $\lambda_0$ before it eventually tends to zero. Since an extended flow equation scheme would couple the forward scattering and backward scattering potentials, this behavior could result in a significant change of the absolute value of $\lambda^B$ in finite-size systems which is found in Ref. \onlinecite{Med02}. We also want to mention that a spatially dependent impurity potential $U(x)$ already couples the forward and backward scattering potentials within the above truncation scheme.\\

If we set the forward scattering potential $\lambda^F$ zero from the beginning and simply neglect the normal ordered contributions of the generated operators $(\Psi_q+h.c.)$, the flow equations close involving only the generator $\eta^{ee}$. With $\lambda^F=0$ for all $\ell$, we obtain the same asymptotic flow equations for $\omega_q$, $u_q$, and $\lambda^B$ and thus the same scenario as above. This simplification will be useful for the evaluation of correlation functions, outlined in the next subsection. In this context, we also want to mention Refs. \onlinecite{Moo93,Egg95,Fen95} where the results of Kane and Fisher are confirmed using Monte Carlo and Bethe ansatz techniques, respectively.  

\subsection{Spectral density at the impurity}
To discuss correlation functions, the observable has to be subjected to the same sequence of infinitesimal transformations that led to the diagonalization of the Hamiltonian. The flow equations for the observable read $\partial_\ell \psi_R(x)=[\eta,\psi_R(x)]$ and differ for $x=0$ and $x\neq0$. At $x=0$, the leading terms are given by
\begin{align}
\psi_R(\ell)=&\psi_R\Big[g+\sum_{q>0}(\varphi_q^R\phi_q^R+\varphi_q^L\phi_q^L)+h\psi_R\dag\psi_L\Big]\notag\\
	+&\psi_L\Big[\bar{g}+\sum_{q>0}(\bar{\varphi}_q^R\phi_q^R+\bar{\varphi}_q^R\phi_q^R)+\bar{h}\psi_L\dag\psi_R\Big]\label{TransObxNULL}
\end{align}
which is symmetric in the left- and right-movers and only the initial condition indicates the evolution of the right-moving field. 
The leading expansion of the field operator at $x\neq0$ is given by
\begin{align}
\psi_R(x,\ell)&=\psi_R(x)\Big[g(x)+\sum_{q>0}\big(\varphi_q^R(x)\phi_q^R(x)\\
&+\varphi_q^L(x)\phi_q^L(x)\big)
+h(x)(\psi_R\dag\psi_L-\psi_L\dag\psi_R)\Big]\quad.\notag
\end{align}
The different expressions are crucial in order to distinguish between the bulk and the boundary regime. 

The flow equations for the field operator turn out to be rather complicated. This is a general feature of the method since flow equations are designed to diagonalize the Hamiltonian and not to yield simple expressions for the observable flow.\cite{Sta02} We will therefore allow for the above mentioned simplification of the Hamiltonian flow by neglecting the forward scattering contribution to the Hamiltonian.  Since now only the generator $\eta^{ee}$ is involved, we can use the calculations of the previous section to determine correlation functions, but we loose the distinction between the bulk and the boundary regime. The truncation scheme of the field operator is thus given by Eq. (\ref{TruncField}) for all $x$. Still, we obtain different expressions for the approximate Green function, as defined in  Eq. (\ref{Green_Trunk_TL}), for attractive and repulsive interaction due to the different fixed point Hamiltonians $H^*$. 

For attractive interaction, the fixed point Hamiltonian $H^*$ is given by the free bosonic bath and the fermionic ground-state $|FDS\rangle$ is obtained with the help of the Kronig relation of Eq. (\ref{Kronig_Relation}). Following the procedure of the last section, we  thus recover that the spectral function is governed by the exact bulk exponent $\alpha=2s^2$ throughout the system even at $x=0$.   

For repulsive interaction, due to the remnant backward scattering potential, the fixed point Hamiltonian has to be transformed further to assure a simple ground-state. First, we again map the non-interacting bosonic bath onto the non-interacting Fermi gas with linear dispersion employing the Kronig relation. The external potential with zero width and zero transmission can now be absorbed by changing the boundary conditions of the physical field operator from periodic to fixed or open boundary conditions. To describe a linearized 1D system with fixed boundary conditions, only right-movers need to be introduced.\cite{Fab95} We thus transform the original ladder-operators of the left- and right-movers according to the following substitution scheme:
\begin{align}
\label{Substitution}
c_k^R\to-ic_k\quad,\quad c_k^L\to ic_{-k}
\end{align}
The sign-change of the momentum in the case of the left-movers expresses the fact that there are only right-movers; the relative phase factor stems from the condition that the physical field operator has to obey fixed boundary conditions, i.e. $\psi(x=0)=\psi_R(x=0)+\psi_L(x=0)=0$.

For finite system size, the transformation is more complicated since a periodic system of size $L$ can only be mapped onto an open system of size $2L$ in order to account for the different wave numbers and to conserve the density of the right-movers.\cite{Fab95} Since we are only interested in results valid in the thermodynamic limit, we will stick to the simple substitution scheme of Eq. (\ref{Substitution}) and divide the transformed fixed point Hamiltonian by two in order to preserve the density of the right-movers, i.e. $H^*\to v_c\sum_kkc_k\dag c_k$.

Performing the substitution also in Eq. (\ref{TruncField}), one sees that only the combination $\tilde{\varphi}_q\equiv(\varphi_q^R+\varphi_q^L)$ enters. With the same approximations as in the previous section and with $\tilde\varphi\equiv n_q^{-1/2}\tilde\varphi_q$ for $q\leq q_c$ and zero otherwise, the spectral density of states at the impurity is given by 
\begin{align}
\rho_R^{1}(\omega)&\approx\frac{\Theta(\tilde\omega)}{2\pi v_c}
\Big[1+\big(2\tilde\varphi^*+\tilde{\varphi}^*{}^2\big)\ln(\tilde\omega/v_cq_c)\Big]\notag\\
&\to\frac{\Theta(\tilde\omega)}{2\pi v_c}\Big(\frac{\tilde\omega}{v_cq_c}\Big)^{\nu-1}\quad,\notag
\end{align}
with $\omega\equiv v_ck_F-\tilde\omega$ and $\nu\equiv(\tilde\varphi^*+1)^2$. As usual, for small interaction algebraic behavior is inferred and the asterisk denotes the limiting value as $\ell\to\infty$.

The differential equation for $\tilde\varphi$ follows from Eq. (\ref{DiffField}) and yields the final result
\begin{align} 
\tilde{\varphi}_q^*\equiv\tilde{\varphi}_q(\ell=\infty)=n_q^{-1/2}(e^{-2E_q^*}-1)\quad.
\end{align}
We thus recover the {\em exact} boundary exponent $\nu=e^{-4E_q^*}\equiv 1/g$, where the last definition corresponds to the notation of Kane and Fisher.

\section{Conclusions}
In the present work, we applied flow equations for Hamiltonians to the TL model with impurity in the limit of weak electron-electron interaction but arbitrary impurity strength. We formulated the TL model by collective density fluctuations which obey Bose statistics, but to represent the impurity potential no use of the bosonization technique was made. We obtained the phase diagram of Kane and Fisher and could provide analytic evidence that it also holds in the regime of intermediate impurity strength. Simplifying the flow of the Hamiltonian by neglecting the forward scattering contribution in the initial Hamiltonian, flow equations for the field operator could be analyzed to yield the exact anomalous dimensions for the different phases as they follow from the bosonization approach. Generalizations of the external one-body potentials are straightforward. The presented approach might also be useful for other systems and resembles an alternative and in a way complementary method compared to functional renormalization schemes.\\ 

The author is grateful to A. Mielke for innumerable discussions on the flow equation method and to V. Meden for comments on the manuscript. The work was supported by the Deutsche Forschungsgemeinschaft.


\begin{thebibliography}{99}
\bibitem{FootBos}
We start from the bosonic representation of 1D interacting fermions mainly for convenience, not for necessity. The flow equations also close within a $1/n$-expansion, see Ref. \onlinecite{Weg94}.
\end{thebibliography}
\end{document}